\documentstyle{article}
\input epsf

\begin{document}

\title{Critical phenomena in gravitational collapse}

\author{Carsten Gundlach \\
Laboratorio de Astrof\'\i sica
Espacial y F\'\i sica Fundamental  \\
Instituto Nacional de Tecnolog\'\i a
Aerospacial \\ PO~Box~50727, 28080~Madrid, Spain}

\date{6 June 1996}

\maketitle


\begin{abstract}

Lecture given at the workshop "Mathematical aspects of theories of
gravitation", Stefan Banach International Mathematical Centre, 7 March
1996. A mini-introduction to critical phenomena in gravitational
collapse is combined with a more detailed discussion of the regularity
of the ``critical spacetimes'' dominating these phenomena.

\end{abstract}


\section{Critical phenomena in gravitational collapse}

Most of the material here is contained also in the paper \cite{PRD}, to
which I refer the reader for details. Here I'll give detail only in the
calculation of the critical exponent, and on how gravity regularizes
self-similar solutions, where I hope it is of more than technical interest.

Initial data for general relativity (GR) may or may not form a black
hole. In a hand-waving way one might compare this to a phase transition,
and there is a ``critical surface'' in superspace (the phase space of GR)
separating the two kinds of initial data. Choptuik \cite{Chop} has explored
this surface in a systematic way. For simplicity he took a massless
scalar field as his matter model, and allowed only spherically symmetric
configurations. (The numerical calculations were still pioneering work,
because they involve length scales spanning many orders of magnitude.) To
explore the infinite-dimensional phase space, he evolved initial data from
a number of one-parameter families of data crossing the critical
surface. Let us call that parameter generically $p$. By a bisection search,
Choptuik found a critical value $p_*$ for each family, such that
data with $p>p_*$ form a black hole, but not data with $p<p_*$. He then
discovered two unforeseen effects:

{\bf Scaling}: Near the critical surface, on the black-hole side of it ($p
> p_*$), where the mass of the black hole final state is small (compared
> to, for example, the ADM mass), it scales as
\begin{equation}
M \simeq C (p-p_*)^\gamma,
\end{equation}
where the overall factor $C$ depends on the family, but the ``critical
exponent'' $\gamma$ is universal between families. For the scalar
field matter, $\gamma \simeq 0.37$. I should stress that this
expression is invariant under redefinitions of $p \to \bar p(p)$ such
that $\bar p(p)$ is differentiable, with $d\bar p/dp\ne 0$ at $p_*$:
To leading order, only $C$ changes under such redefinitions. To
clarify this, and to avoid speaking in terms of one-parameter families
altogether, one can formally introduce coordinates $(x_0,x_i)$ on
superspace such that the coordinate surface $x_0=0$ is the critical
surface. Then, for small positive $x_0$, the black hole mass would be of
the form
\begin{equation}
M \simeq f(x_i)\,x_0^{\gamma},
\end{equation}
and again this form is invariant under coordinate diffeomorphism-s such that
$x_0=0$ remains the critical surface.

{\bf Universality}: Near the critical surface, on either side of it ($p
\simeq p_*$ ), initial data evolve towards an intermediate asymptotic
solution (which I'll call $Z_*$), which is again universal with respect to
initial data. (Clearly this solution cannot be a full-blown attractor,
because all data must veer off it eventually towards forming either a black
hole or towards dispersion. But it is an attractor of co-dimension one, or
intermediate asymptotic.) This solution shows self-similarity. In the case
of scalar field matter, this self-similarity is discrete, showing up as an
``{\bf echoing}'' in the logarithm of spacetime scale, with period $\Delta
\simeq 3.44$.

The matter model of Choptuik is special in that, even coupled to general
relativity, it has no intrinsic scale. (This is equivalent to saying that
in geometric units, $c=G=1$, the action has no dimensionful parameters.)
Therefore, from dimensional analysis alone, there can no be no static,
star-like solutions, and hence no minimum black hole mass. Generic matter does have
both an intrinsic scale (or several), and star-like
solutions. Astrophysical black holes thus probably have a minimum mass
given by the Chandrasekhar mass. To make infinitesimally small black holes
from ordinary matter, one would have to use initial conditions (for example
rapid implosion) not arising in astrophysics. The important point is that
for any matter there is at least {\bf some} region of superspace where
critical phenomena arise.  Nevertheless, the interest of critical phenomena
does not lie in astrophysics, but in the dynamics of GR (with or without
matter). Recent work by Choptuik, Bizon and Chmaj \cite{EYM} (see Piotr
Bizon's talk, this volume) has clarified the role of matter
scales. Investigating the spherical collapse of Einstein-Yang-Mills, they
find two regimes: one with the usual critical phenomena dominated by a
self-similar intermediate attractor (``second-order phase transition''), and
one dominated by the Bartnik-McKinnon solution, which is a finite mass,
static intermediate attractor, and therefore having a mass gap
(``first-order phase transition'').

Critical phenomena have also been found in two one-parameter families of
matter models in spherical symmetry (not to be confused with one-parameter
families of data!): The first family is that of the perfect fluids with
$p=k\rho$, $k$ a constant \cite{EvCol,Koike,Maison}. The other family is
that of the constant-curvature, two-dimensional sigma models, characterised
by the dimensionless curvature parameter $\kappa$ \cite{HE3}. (This family
contains the free complex scalar field \cite{HE1}, an inflaton-dilaton
model \cite{ChopLieb}, and an axion-dilaton model \cite{EHH} as special
cases.) It is known for the latter family, and likely for the
second, that at some value of the parameter $\kappa$ (and perhaps $k$) the
critical solution switches over from continuous to discrete
self-similarity. Moreover, the critical exponent $\gamma$ depends on the
parameter $\kappa$ or $k$.

Historically, the second occurrence of critical phenomena was found in the
collapse of axisymmetric gravitational waves (time-symmetric initial data,
and hence with zero angular momentum) \cite{AbrEv}. Because of the much
greater numerical difficulty, there are fewer experimental data, and
regrettably, there has been no follow-up work so far. Axisymmetric waves
are highly interesting in two aspects: They go beyond spherical symmetry,
and they are vacuum data. It would also be extremely interesting to see
what happens for initial data with angular momentum, because black holes
can of course have angular momentum, but it must be smaller than the mass.

\section{Self-similarity in GR}

The critical solution for each matter model has the property of being
self-similar. With hindsight one can see that this is the form
scale-invariance takes in the problem. Self-similarity however takes the
novel form of echoing, or discrete self-similarity, in some models, notably
Choptuik's scalar field and axisymmetric gravitational waves.

The concept of (continuous) self-similarity (CSS) (or homotheticity)
has been defined in a relativistic context \cite{CahTaub} as
the presence of a vector field $\chi$ such that
\begin{equation}
{\cal L}_\chi g_{ab} = 2 g_{ab},
\end{equation}
where ${\cal L}_\chi$ denotes the Lie derivative. I now introduce the concept
of discrete self-similarity (DSS). In this symmetry there exist a
diffeomorphism $\phi$ and a real constant $\Delta$ such that, for any
integer $n$,
\begin{equation} 
\label{discrete}
\left(\phi_*\right)^n g_{ab} = e^{2n\Delta} g_{ab},
\end{equation}
where $\phi_*$ is the pull-back of $\phi$.

To see what DSS looks like in coordinate terms, we introduce coordinates
$(\tau,x^\alpha)$, such that if a point $p$ has coordinates
$(\tau,x^\alpha)$, its image $\phi(p)$ has coordinates
$(\tau+\Delta,x^\alpha)$. One can verify that DSS in these
coordinates is equivalent to
\begin{equation}
\label{rescaled}
g_{\mu\nu}(\tau,x^\alpha) = e^{2\tau} \tilde
g_{\mu\nu}(\tau,x^\alpha),
\quad \hbox{where} \quad \tilde
g_{\mu\nu}(\tau,x^\alpha) = \tilde
g_{\mu\nu}(\tau+\Delta,x^\alpha)
\end{equation}
In other words, the DSS acts as a discrete isomorphism on the rescaled
metric $\tilde g_{\mu\nu}$. $\tau$ is intuitively speaking the logarithm
of spacetime scale. 

In order to clarify the connection between CSS and DSS, one may define
a vector field $\chi\equiv \partial / \partial \tau$, although there
is no unique $\chi$ associated with a given $\phi$. The discrete
diffeomorphism $\phi$ is then realized as the Lie dragging along
$\chi$ by a distance $\Delta$. Clearly, CSS corresponds to DSS for
infinitesimally small $\Delta$, and hence for all $\Delta$, and is in
this sense a degenerate case of DSS. In this limit, $\chi$ becomes
unique. 

\section{Universality and the critical solution}

The critical solution dominating critical phenomena for a given matter
model (and perhaps choice of symmetry, such as spherical symmetry), has two
essential properties. First, it must be self-similar, either CSS or
DSS. Secondly, it must have exactly one unstable mode. There are examples
of self-similar solutions with more than one unstable mode, with exactly
one (i.e. genuine critical solutions), but not with none: the latter would
constitute a violation of cosmic censorship in the strongest possible
sense, that of a naked singularity arising from generic initial data. The
presence of exactly one unstable mode, on the other hand, constitutes a
dynamical explanation of the universality of the critical exponent. In this
explanation, near-critical data $p\simeq p_*$ (data near the critical
surface) are precisely those in which the one unstable mode is initially
small. Skimming along the critical surface, they are attracted towards the
critical solution, which is either a fixed point (CSS) or a limit cycle
(DSS), until the growing mode eventually takes over and ejects the
trajectory, either towards black hole formation or towards dispersion, in a
unique manner. The solution has forgotten from which initial data it came,
up to one parameter (roughly speaking, the number of echos, or time spent
on the intermediate asymptotic) which will eventually determine the black
hole mass. Figure 1 illustrates this behaviour for DSS.

A given spacetime does not correspond to a unique trajectory in superspace,
because it can be sliced in different ways. A point in superspace does
correspond to a unique spacetime, but not to a unique trajectory. In other
words, there is no preferred time-evolution flow on superspace, and
therefore Figure 1 is true only for a fixed slicing condition, something
extraneous to GR. I am currently trying to understand this issue. A
possible solution is the fact that the preferred time evolution (choice of
lapse and shift) should correspond to a ``renormalisation group flow'', or
change of scale. This is the case for example if one takes $\tau$ as
defined above as the time variable. The analogy with critical phenomena in
statistical mechanics would then be deeper: superspace corresponds to the
space of Hamiltonians, and and the preferred time evolution to the
renormalisation group flow.

Constructing the critical solution, one has to proceed in two
steps. Imposing certain regularity conditions and self-similarity, one
obtains a non-linear eigenvalue problem, which may have a solution. As a
second step one has to calculate the spectrum of its linear perturbations
and check that there is exactly one growing mode. 

In the following I restrict myself to spherical symmetry.  The regularity
conditions just mentioned are imposed in two places. One is the centre of
spherical symmetry. There one imposes the absence of a conical singularity
in the 3-metric, and that all fields be either even or odd in $r$, depending
on their being vector or scalar under rotations. The other set of
regularity conditions needs to be discussed in some more detail.

The general spherically symmetric solution of the wave equation in flat
space is 
\begin{equation}
\phi(r,t) = r^{-1} \, [f(r-t)+g(r+t)],
\end{equation}
with $f$ and $g$ arbitrary functions. Furthermore, $\phi$ is DSS (in flat
space) if $\phi(r,t) = \phi(e^\Delta r,e^\Delta t)$ for some $\Delta$.  One
easily derives that the general DSS solution in flat space is of the form
\begin{equation}
\phi(r,t) = (1+z^{-1})F[\tau + \ln(1+z)]
+ (1-z^{-1})G[\tau + \ln(1-z)], \quad \tau\equiv \ln t,
\quad z\equiv r/t,
\end{equation}
where $F$ and $G$ are now periodic (with period $\Delta$), but otherwise
arbitrary functions.  

With the exception of $\phi=0$, a DSS solution can only be regular at
either $r=0$ ($F=G$), or at $r=t$ ($G=0$), or at $r=-t$ ($F=0$).  All DSS
solutions (except the zero solution) are singular at the point
$(r=0,t=0)$. Coupling the wave equation to GR in spherical symmetry changes
the dynamics, but not the degrees of freedom. (There are no spherically
symmetric gravitational waves.)

Surprisingly, the presence of gravity acts as a regulator. In the presence
of gravity, there is (at least) one DSS solution which is regular at both
$r=0$ and at the past light cone of $(r=0,t=0)$ (the generalisation of
$r=-t$ to curved spacetime) in the sense of being analytic, and regular, in
a weaker sense, also at the future light cone. $(r=0,t=0)$ remains
singular, in the sense of a curvature singularity. This solution is
precisely the intermediate attractor dominating near-critical collapse. It
is found by numerically solving a non-linear hyperbolic eigenvalue problem,
with periodicity in some coordinate $\tau$, and regularity now imposed at
both $r=0$ and the past light cone. For all details I refer the reader to
\cite{PRL,PRD}. I'll come back to the regulating effect of gravity in a
moment, but first I want to conclude my mini-review with an idea of how one
calculates the critical exponent.

\section{Calculation of the critical exponent}
 
The calculation of the critical exponent is such a nice piece of
dimensional analysis \cite{EvCol,Koike,HE2} that I simply must sketch it
here. For simplicity of notation I assume CSS of the critical solution, the
more generic DSS case is given in the paper \cite{PRD}. Mathematically, the
calculation that follows (for CSS) is identical with that of the critical
exponent governing the correlation length near the critical point in
statistical mechanics \cite{Yeomans}.

Let $Z$ stand for the variables of the problem in a first-order
formulation, in spacetime coordinates and matter variables adapted to the
problem \cite{PRL,PRD}. $Z(r)$ is an element of the phase space, and
$Z(r,t)$ a solution.  The self-similar solution is of the form
$Z(r,t)=Z_*(r/t)$. (I won't spell out what $Z$ stands for or how $r$ and
$t$ are defined.) In the echoing region, where $Z_*$ dominates, we
linearise around it. To linear order, the solution must be of the form
\begin{equation}
Z(r,t) \simeq Z_*\left(r\over t\right) + \sum_{i=1}^\infty C_i(p)
(-t)^{\lambda_i} \delta_i Z\left(r\over t\right).
\end{equation} 
Here, the general form of the linear perturbations follows from the form of
the background solution $Z_*$. Their coefficients $C_i$ depend in a
complicated way on the initial data, and hence on $p$. If $Z_*$ is a
critical solution, by definition there is exactly one $\lambda_i$ with
negative real part (in fact it is purely real), say $\lambda_1$. As $t\to
0$, all other perturbations vanish, and in the following we consider this
limit, and retain only the perturbation with $i=1$. Furthermore, by
definition the critical solution corresponds to $p=p_*$, so we must have
$C_1(p_*)=0$. Linearising around $p_*$, we obtain
\begin{equation}
\lim_{t\to 0} Z(r,t) \simeq Z_*\left(r\over t\right) + {dC_1\over dp} (p-p_*)
(-t)^{\lambda_1} \delta_1 Z\left(r\over t\right).
\end{equation}
This form holds over a range of $t$, that is, is an approximate
solution. Now we extract Cauchy data by picking one particular value of $t$
within that range, namely $t_p$ defined by
\begin{equation}
{dC_1\over dp} (p-p_*)
(-t_p)^{\lambda_1} \equiv \epsilon,
\end{equation}
where $\epsilon$ is some constant $\ll 1$, so that at $t_p$ the linear
approximation is still valid. (The suffix $p$ indicates that $t_p$ depends
on $p$.) At sufficiently small $t$, the linear perturbation $Z_1$ has grown
so that the linear approximation breaks down. Later on a black hole forms. 
The crucial point is that we need not follow this evolution in detail. It
is sufficient to note that the Cauchy data at $t=t_p$ depend on $r$ only in
the combination $r/t_p$, namely
\begin{equation}
Z(r,t_p) \simeq Z_*\left(r\over t_p\right) + \epsilon \ \delta_1
Z\left(r\over t_p\right).
\end{equation}
As furthermore the field equations do not have an intrinsic scale, it
follows that the solution based on those data must be of the form
\begin{equation}
Z(r,t) = f\left({r\over t_p}, {t-t_p\over t_p}\right) 
\end{equation}
throughout, even when the black hole forms and perturbation theory breaks
down, and still after it has settled down
and the solution no longer depends on $t$. (This solution holds only for
$t>t_p$, because in its initial data we have neglected the perturbation
modes with $i>1$, which are growing, not decaying, towards the past.)
Because the black hole mass has dimension length, it must be proportional
to $t_p$, the only length scale in the solution,
\begin{equation}
M \propto t_p \propto (p-p_*)^{-{1\over \lambda_1}},
\end{equation}
and we have found the critical exponent. 

For completeness I mention that the scaling law is modified when the
critical solution is DSS. On the straight line relating $\ln M$ to
$\ln(p-p_*)$ a periodic wiggle of small amplitude is superimposed. This
wiggle is again universal with respect to families of initial data, and
there is only one free parameter for each family to be adjusted,
corresponding to a shift of the wiggly line in the $\ln(p-p_*)$
direction. (No separate shift in the $\ln M$ direction is required.)

\section{Gravity as a regularizer of self-similar solutions}

I now come back to the critical solution, which by definition is DSS and
regular. As I said before, DSS is imposed as periodicity in a coordinate
system of the form (\ref{rescaled}) adapted to the problem, and boundary
conditions at $r=0$ arise in a straightforward manner, from the necessity
of avoiding a conical singularity. The generalisation of $t=-r$ to curved
spacetime is more interesting. 

From the form (\ref{rescaled}) of the metric it is easy to see that the
curvature blows up as $\tau\to -\infty$. Furthermore, this singularity is a
``point'' in, for example, the following sense. Let
$(\tau_1,\zeta_1,\theta_1,\varphi_1)$ and
$(\tau_2,\zeta_2,\theta_2,\varphi_2)$ be two points. Their geodesic
distance vanishes as $e^\tau$ as $\tau_1\to\tau_2\to -\infty$, for any
values of $(\zeta_1,\theta_1,\varphi_1)$ and
$(\zeta_2,\theta_2,\varphi_2)$.

The generalisation of $r=t$ to curved spacetime is the past light cone of
this singularity. We use the remaining freedom in the coordinate system
(\ref{rescaled}) to label this light cone $\zeta=0$. From the analogy with
the general solution of the wave equation in flat space one would assume
that any solution regular at $r=0$ is singular here, showing an infinite
number of oscillations of the form $\phi\sim F(\ln \zeta)$ (with $F$
periodic) as $\zeta\to 0$.

But this is not so, and I now show why. In flat space, the inward and
outward travelling modes are $\phi' + \dot\phi$ and $\phi' - \dot\phi$. Let
us call their curved-space equivalents $X_+$ and $X_-$. It is $X_-$ that we
expect to be singular at $\zeta=0$. The equation for $X_-$, to leading
order in $\zeta$, is
\begin{equation}
\label{approx}
X_{-,\zeta} = {A(\tau) X_- - X_{-,\tau} + C(\tau) \over \zeta B(\tau)},
\end{equation}
where the coefficients $A$, $B$ and $C$ are constructed from the other
fields and are therefore periodic in $\tau$. 

This approximate equation admits an exact
general solution, namely
\begin{equation}
X_- = X_-^{{\rm inhom}}(\tau) + X_-^{{\rm hom}}(\zeta,\tau).
\end{equation}
The particular inhomogeneous solution $X_-^{{\rm inhom}}$ is defined
as the unique solution of
\begin{equation}
A X_-^{{\rm inhom}} - X_{,\tau}^{{\rm inhom}} + C = 0
\end{equation}
with periodic boundary conditions. This solution exists and is unique,
unless the average value of $A$ vanishes. The general homogeneous
solution $X_-^{{\rm hom}}$ is of the form
\begin{equation}
 X_-^{{\rm hom}} = \zeta^{A_0\over B_0} \ e^{\int A \ - {A_0\over
B_0} \int B} \ F\left[\tau + {\int B - \ln |\zeta| \over B_0}\right].
\end{equation}
where $A_0$ is the average value of the periodic function $A$, and $\int A$
is its principal function {\it after the average value has been
subtracted}, so that $\int A$ is by definition also periodic. The periodic
function $F$ depends on the initial data for the equation (\ref{approx}).
I have not given the expressions for $A$, $B$, and $C$ here, but in flat
space $A$ and $B$ vanish. In the critical solution, which is far from flat,
$A_0$ is negative and $B_0$ is positive. The exponent of $\zeta$,
$A_0/B_0$, is negative for the critical or neighbouring solutions. In
consequence, $X_-^{{\rm hom}}$ only has two alternatives, it either blows
up at $\zeta=0$, or it is analytic there (for $F\equiv 0$). To impose
regularity at $\zeta=0$, it suffices therefore to impose
\begin{equation}
\label{condition}
A(\tau) X_- - X_{-,\tau} + C(\tau) = 0
\end{equation}
at $\zeta=0$. As it happens, there is locally just one solution which obeys
this condition, as well as the other boundary conditions.  

As I have said, $A$ vanishes in flat space. But then the equation
(\ref{condition}) has no solution $X_-$ {\it with periodic boundary
conditions}, because the average value of $C$ does not vanish. But the
presence of the term proportional to $X_-$ changes the character of the
equation quantitatively, and a solution always exists. (In the limit as $A$
vanishes, this solution blows up.)

We have now found the solution in the past light cone of the
singularity. The data on the past light cone then determine the solution up
to the future light cone. (We can go no further because the future light
cone of the singularity is a Cauchy horizon.)  Calculating this maximal
extension numerically requires two more nontrivial changes of coordinate
system, giving rise to fresh eigenvalue problems. We arrange the final
coordinate patch so that once more the future light cone is a coordinate
line. Now it is $X_+$ which is potentially singular. Its equation is of the
same form (\ref{approx}), where $X_+$ replaces $X_-$, and $\zeta$ has been
redefined so that $\zeta=0$ is now the future light cone. Now, however,
there is no freedom left to adjust any data in order to set $F\equiv 0$,
and in fact $F$ does not vanish with the data we have in hand. So the
solution cannot be analytic at the future light cone. In contrast to the
past light cone, however, both $A_0$ and $B_0$ are positive. This means
that $X_+^{{\rm hom}}$, with an infinite number of oscillations as
$\zeta=0$ is approached, is present, but vanishes at $\zeta=0$ as a (small)
positive power of $\zeta$. $X_+$ exists at $\zeta=0$, but $\partial
X_+/\partial \zeta$ does not. From the Einstein equations, which I have not
given here, it follows that the metric and all its first derivatives exist,
but not some of its second derivatives. Nevertheless these particular
second derivatives cancel out of all components of the Riemann tensor, so
that the Riemann tensor exists (but not some of its first derivatives).

Figure 2 summarises the global situation, with one angular coordinate
suppressed. We have spherically symmetric, discretely self-similar
spacetime, with a single point-like singularity. The self-similarity
corresponds roughly speaking to periodicity in the logarithm of the
distance from the singularity. The past light cone of the singularity is
totally regular, in no way distinguished from other spacetime points. The
future light cone, or Cauchy horizon, is perhaps as regular as one can
expect, with the scalar matter field $C^0$, the metric $C^1$ and the
Riemann tensor $C^0$. In particular it carries well-defined null data
(which are of course self-similar), and there exists a regular,
self-similar (and of course, non-unique) extension of the spacetime inside
the Cauchy horizon, with only the horizon itself of limited
differentiability. Surprisingly, the null data on the horizon are very
small, so that the extension can be made almost flat and empty. The
situation is similar in the other two cases where the critical solution has
been calculated up to the Cauchy horizon \cite{HE1,EHH}. The spacetime
there is CSS, but also almost flat at the horizon. A limiting case arises
in the closed-form solution of Roberts \cite{Roberts}, where the null data
on the horizon vanish exactly, so that a possible extension inside the
light cone is flat empty space. (In the Roberts solution, the past light
cone also carries zero data, in contrast to the other examples, where the
past light cone is very far from flat.) Finally I should add that at large
spacelike distances from the singularity, spacetime becomes asymptotically
conical, with a constant defect solid angle.

Of course, all these global considerations are not relevant to critical
collapse, where the spacetime asymptotes the critical spacetime in some
bounded region inside the past light cone. Therefore, neither the
singularity nor the asymptotically conical region, nor the Cauchy horizon
appear. 

\section{Conclusions}

Looking back on what has become a small industry in the past two years, I
think that the dynamical mechanism of universality, scaling and echoing
is now understood in an intuitive way. Furthermore we can calculate the
echoing period $\Delta$ and critical exponent $\gamma$ as nonlinear
eigenvalues, in a manner distinct from that of fine-tuning data in numerical
experiment. 

Critical phenomena have also given new material to the study of cosmic
censorship. There are self-similar spacetimes (spherically symmetric,
except for one axisymmetric example) with a naked singularity that have an
infinity of decaying linear perturbation modes opposed to only one
increasing perturbation mode. This still implies cosmic censorship in the
sense that a generic perturbation, which will contain some small fraction
of that unstable mode, destroys the naked singularity (either by forming a
horizon or by avoiding a singularity altogether, depending on the sign of
the mode amplitude). But we are very close to a violation, in that we only
have to set one mode out of an infinity of modes to zero to get the naked
singularity. A single generic parameter in the initial data provides a
sufficient handle to do this. (In the notation above, we only have to set
the one amplitude $C_1$ equal to zero, and if $C_1$ depends in some way on
$p$, we can do this by adjusting $p$.) Therefore by arbitrary fine-tuning
of one generic parameter in the initial data one can obtain asymptotically
flat spacetimes in which a region of arbitrarily large curvature is visible
to an observer at infinity.

There are a number of important open questions. Are there additional
unstable perturbations of the critical solution among the nonspherical
modes? Are there critical solutions in axisymmetry, or even lower symmetry?
Is DSS more generic than CSS? How do the charge and angular momentum of the
black hole scale as one fine-tunes initial data with charge and angular
momentum, with the aim of making a black hole of infinitesimal mass (and
not caring about charge and angular momentum)? Is there a preferred flow on
superspace which would complete the analogy with the renormalisation group
flow in statistical mechanics? Is there even statistical physics hidden
somewhere? (Very, very unlikely, but who would have thought it of black
holes pre-1974?) Can someone give an existence proof for the critical
solutions, which so far have only been constructed numerically? Can one
prove that they must have the limited differentiability (metric $C^1$ etc.)
at the Cauchy horizon which I have described, not more or less?


\begin{figure}
\epsfysize=14cm
\centerline{\epsffile{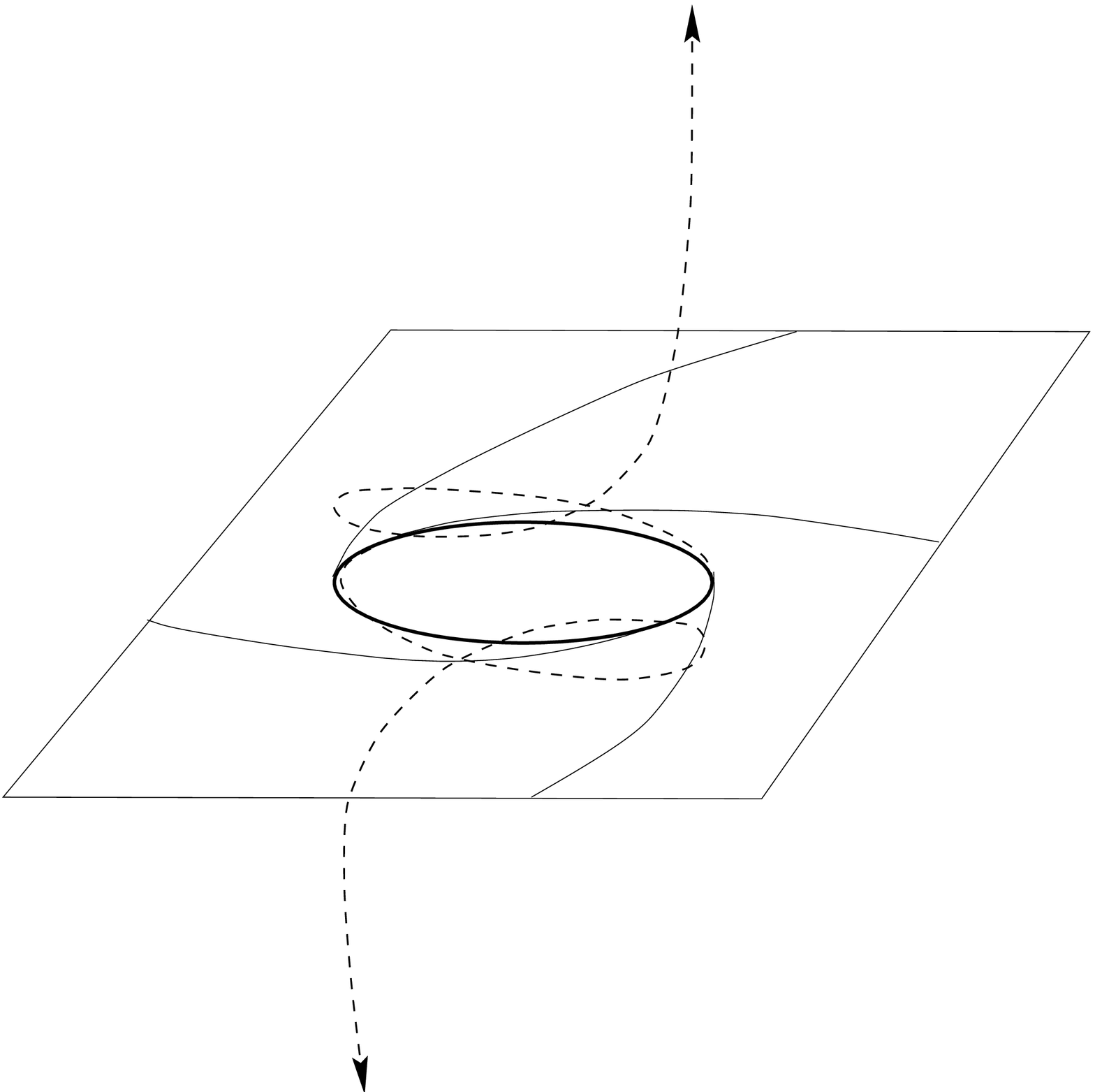}}
\caption{Figure 1: The phase space picture for discrete
self-similarity. The plane represents the critical surface. (In reality
this is a hypersurface of co-dimension one in an infinite-dimensional
space.) The circle (fat unbroken line) is the limit cycle representing the
critical solution. The thin unbroken curves are spacetimes attracted to
it. The dashed curves are spacetimes repelled from it. There are two
families of such curves, labeled by one periodic parameter, one forming a
black hole, the other dispersing to infinity. Only one member of each
family is shown.}
\end{figure}

\begin{figure}
\epsfysize=16cm
\centerline{\epsffile{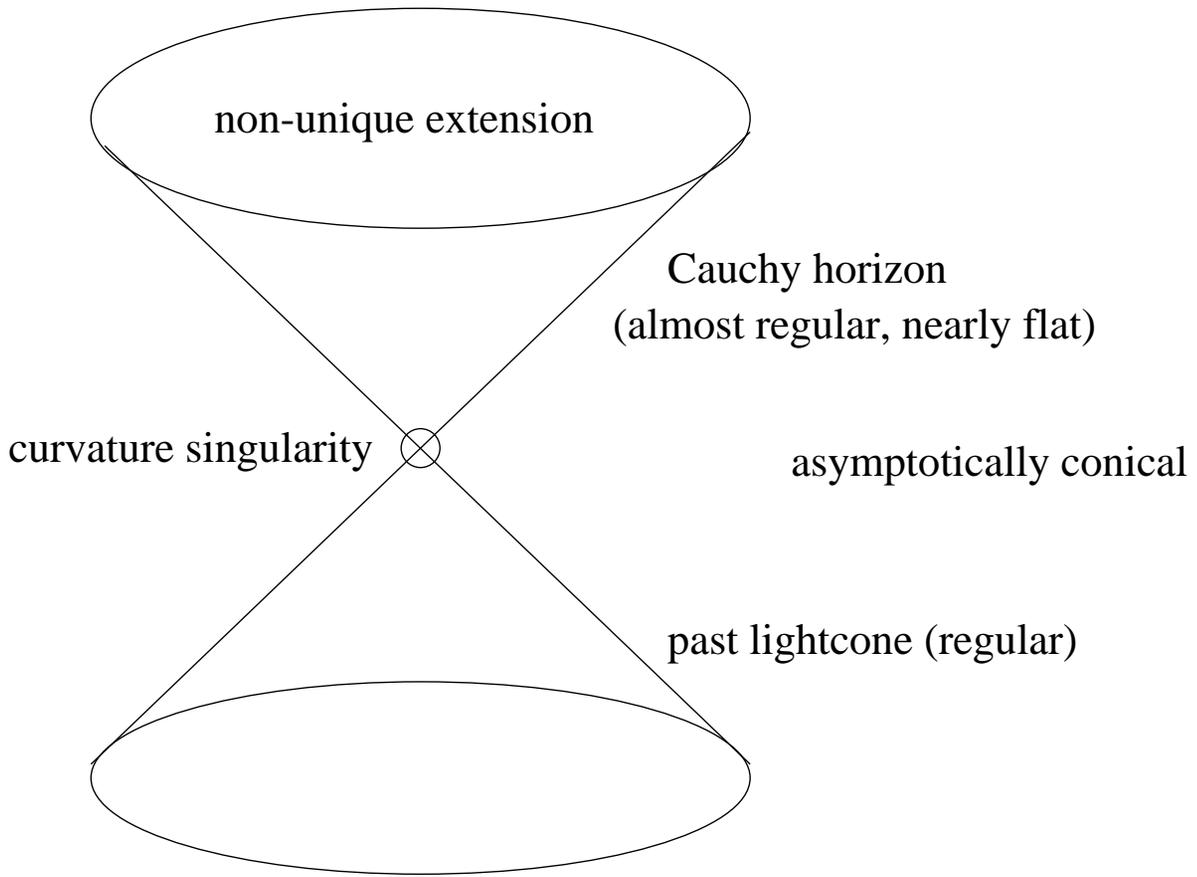}}
\caption{Figure 2: The global structure of the critical spacetimes. One
dimension in spherical symmetry has been suppressed.}
\end{figure}


\end{document}